\documentclass[
reprint,           
superscriptaddress,
amsmath,           
amssymb,           
aps,               
prd,               
notitlepage,       
floatfix,          
nofootinbib 
]{revtex4-1}

\usepackage{tensor}     
\usepackage{graphicx}   
\usepackage[
colorlinks=true,        
citecolor=blue,         
linkcolor=blue,         
urlcolor=blue           
]{hyperref}             
\usepackage{bm}         
\usepackage{xcolor}     
\usepackage{tabulary}
\usepackage{color}      
\usepackage[utf8]{inputenc} 
\usepackage[section]{placeins} 
\usepackage{ulem}
\newcommand{\nc}{\newcommand*} 

\nc{\al}{\alpha}
\nc{\s}{\sigma}
\nc{\dt}{\delta}
\nc{\Dt}{\Delta}
\nc{\Ld}{\Lambda}
\nc{\p}{\partial}
\nc{\om}{\omega}
\nc{\Om}{\Omega}
\nc{\rd}{\mathrm{d}}
\nc{\Od}[1]{\mathcal{O}(#1)} 
\nc{\kp}{\kappa}
\nc{\one}{\uppercase\expandafter{\romannumeral1}}
\nc{\two}{\uppercase\expandafter{\romannumeral2}}
\nc{\three}{\uppercase\expandafter{\romannumeral3}}
\def\({\left(}
\def\){\right)}
\def\[{\left[}
\def\]{\right]}
\def\e{\begin{equation}}
\def\q{\end{equation}}
\def\m{\begin{eqnarray}}
\def\n{\end{eqnarray}}
\nc{\Eq}[1]{Eq.~\eqref{#1}}     
\nc{\Fig}[1]{Fig.~\ref{#1}}     
\nc{\Table}[1]{Table~\ref{#1}}  
\nc{\Sec}[1]{Sec.~\ref{#1}}     
\nc{\Msun}{M_\odot}             
\nc{\fpbh}{f_{\mathrm{pbh}}}    
\nc{\fpbhn}{f_{\mathrm{pbh0}}}    
\nc{\mR}{\mathcal{R}} 
\nc{\seq}{\sigma_{\mathrm{eq}}}
\nc{\ogw}{\Omega_{\mathrm{GW}}}
\nc{\gpcyr}{\mathrm{Gpc}^{-3}\,\mathrm{yr}^{-1}}
\nc{\lvc}{LIGO/Virgo} 
\nc{\SNR}{\mathrm{SNR}} 
\nc{\mmin}{{m_{\mathrm{min}}}}
\nc{\mmax}{{m_{\mathrm{max}}}}
\nc{\Mmin}{{M_{\mathrm{min}}}}
\nc{\fmin}{{f_{\mathrm{min}}}}
\nc{\VT}{\mathrm{VT}}
\nc{\rhoGW}{\rho_{\mathrm{GW}}}
\nc{\vth}{\vec{\theta}}
\nc{\vd}{\vec{d}}
\nc{\vla}{\vec{\lambda}}
\nc{\Nobs}{N_{\mathrm{obs}}}
\nc{\av}[1]{\langle #1 \rangle} 
\nc{\km}{\mathrm{km}}
\nc{\Mpc}{\mathrm{Mpc}}
\nc{\Tobs}{T_{\mathrm{obs}}}
\nc{\Ntemp}{N_{\mathrm{temp}}}
\nc{\addref}{[\textcolor{red}{add ref}] } 
\nc{\eg}{\textit{e.g.~}}
\nc{\app}{\approx}
\nc{\hf}{\frac{1}{2}}
\nc{\discuss}{\textcolor{red}{Add discussion here!}}
\nc{\red}[1]{\textcolor{red}{#1}}
\nc{\mH}{\mathcal{H}}
\nc{\cs}{c_s^2}
\nc{\Sij}[1]{S_{ij}^{(#1)}}
\nc{\vi}[1]{v_i^{(#1)}}
\nc{\no}{\nonumber}
\def\<{\left\langle}
\def\>{\right\rangle}

\nc{\bk}{\bm{k}}
\nc{\bq}{\bm{q}}
\nc{\bp}{\bm{p}}
\nc{\bl}{\bm{l}}
\nc{\bx}{\bm{x}}
\nc{\be}{\mathbf{e}}
\nc{\mS}{\mathcal{S}}
\nc{\te}{\tilde{\eta}}
\nc{\tp}{\tilde{p}}
\nc{\tk}{\tilde{k}}
\nc{\tx}{\tilde{x}}
\nc{\tF}{\tilde{F}}
\nc{\tA}{\tilde{A}}
\nc{\mkpq}{|\bk-\bp-\bq|}
\nc{\mpq}{|\bp-\bq|}
\nc{\mkp}{|\bk-\bp|}
\nc{\mSi}[1]{\mS^{(#1)}({\bk, \eta})}
\nc{\vk}{\vec{k}}
\nc{\kstar}{k_*}
\nc{\xstar}{x_*}
\nc{\mpbh}{m_{\rm{pbh}}}
\nc{\Ci}{\mathrm{Ci}}
\nc{\Si}{\mathrm{Si}}
\nc{\fnl}{f_\mathrm{NL}}
\nc{\gnl}{g_\mathrm{NL}}
\nc{\Fnl}{F_\mathrm{NL}}
\nc{\Gnl}{G_\mathrm{NL}}
\nc{\togw}{\tilde{\Omega}}
\nc{\md}{\mathrm{d}^3}
\nc{\taugw}{\tau_{\mathrm{GW}}}
\nc{\tauinst}{\tau_{\mathrm{inst}}}
\nc{\MSmax}{M_{S}^{\mathrm{max}}}

\renewcommand{\vec}[1]{\boldsymbol{#1}} 

\begin{document}
	
\title{Probing ultralight dark matter with future ground-based gravitational-wave detectors}
\author{Chen Yuan}
\email{yuanchen@itp.ac.cn}
\affiliation{CAS Key Laboratory of Theoretical Physics, 
	Institute of Theoretical Physics, Chinese Academy of Sciences,
	Beijing 100190, China}
\affiliation{School of Physical Sciences, 
	University of Chinese Academy of Sciences, 
	No. 19A Yuquan Road, Beijing 100049, China}

\author{Richard Brito}
\affiliation{Dipartimento di Fisica, ``Sapienza" Università di Roma \& Sezione INFN Roma1, Piazzale Aldo Moro 5, 
00185, Roma, Italy}

\author{Vitor Cardoso}
\affiliation{CENTRA, Departamento de F\'{\i}sica, Instituto Superior T\'ecnico -- IST, Universidade de Lisboa -- UL,
	Avenida Rovisco Pais 1, 1049 Lisboa, Portugal}

\date{\today}
	
\begin{abstract}
Ultralight bosons are possible fundamental building blocks of nature, and promising dark matter candidates. They can trigger superradiant instabilities of spinning black holes (BHs) and form long-lived ``bosonic clouds''  that slowly dissipate energy through the emission of gravitational waves (GWs). Previous studies constrained ultralight bosons by searching for the stochastic gravitational wave background (SGWB) emitted by these sources in LIGO data, focusing on the most unstable dipolar and quadrupolar modes. Here we focus on scalar bosons and extend previous works by: (i) studying in detail the impact of higher modes in the SGWB; (ii) exploring the potential of future proposed ground-based GW detectors, such as the Neutron Star Extreme Matter Observatory, the Einstein Telescope and Cosmic Explorer, to detect this SGWB. We find that higher modes largely dominate the SGWB for bosons with masses $\gtrsim 10^{-12}$ eV, which is particularly relevant for future GW detectors. By estimating the signal-to-noise ratio of this SGWB, due to both stellar-origin BHs and from a hypothetical population of primordial BHs, we find that future ground-based GW detectors could observe or constrain bosons in the mass range $\sim [7\times 10^{-14}, 2\times 10^{-11}]$ eV and significantly improve on current and future constraints imposed by LIGO and Virgo observations. 
\end{abstract}
	
	
\maketitle
	
	
\section{Introduction}
The detection of gravitational waves (GWs) from the mergers of binary black holes (BHs) and binary neutron stars~\cite{LIGOScientific:2018mvr,Abbott:2020niy} has marked the dawn of GW astronomy. GWs not only provide a powerful tool to test fundamental physics in the strong-field regime \cite{Barack:2018yly,Baibhav:2019rsa,LIGOScientific:2019fpa,Abbott:2020jks,Sathyaprakash:2019yqt,Barausse:2020rsu}, but also offer a unique opportunity to hunt for dark matter (DM)~\cite{Barack:2018yly,Baibhav:2019rsa,Bertone:2019irm}. 
Of all possible DM candidates, ultralight bosons~\cite{Preskill:1982cy,Abbott:1982af,Dine:1982ah,Arvanitaki:2009fg,Arvanitaki:2010sy,Essig:2013lka,Brito:2015oca,Marsh:2015xka,Hui:2016ltb,Annulli:2020lyc,Chadha-Day:2021szb} are among the most promising to lead to observational signatures in the GW spectrum. For bosons with masses $m_s c^2\sim [10^{-20}, 10^{-10}]$ eV their typical oscillation frequency $f_s\equiv m_s c^2/h\sim  [10^{-6}, 10^{4}]$ Hz falls right within the frequency band of current and future GW detectors whereas their (reduced) Compton wavelength $\lambdabar_s \equiv \hbar/(m_s c)\sim  [1, 10^{10}]\, G M_{\odot}/c^2$ matches the typical size of astrophysical BHs. This leads to the exciting possibility that signatures arising from the interaction of BHs with bosonic fields might be observable with GW observations~
\cite{Brito:2015oca,Barack:2018yly}.

The most interesting of those signatures arise from the fact that massive bosons can render spinning BHs unstable against superradiant extraction of energy and angular momentum from the BH.
Take a spinning BH with horizon angular velocity $\Omega_{\mathrm{H}}$. A field with azimuthal dependence $e^{im\varphi}$, with $\varphi$ the azimuthal angle, is unstable when its frequency $\omega_R\sim \mu  \equiv m_s/\hbar$ (from now on we use units $G=c=1$) satisfies the condition $0<\omega_R<m\Omega_{\mathrm{H}}$ \cite{Press:1972zz,Detweiler:1980uk,Cardoso:2004nk,Dolan:2007mj,Brito:2015oca}. 
Due to the superradiant instability, the BH spins down, transferring its energy and angular momentum to the bosonic field until reaching the saturation point where $\omega_{R}\sim \mu \sim m\Omega_{\mathrm{H}}$~\cite{Brito:2014wla,East:2017mrj,East:2018glu}. This leads to the formation of a macroscopic, time-varying and non-axisymmetric, rotating boson cloud, which on very long timescales dissipates its energy through the emission of nearly monochromatic GWs with frequency $f_0=\omega_R/\pi\approx \mu/\pi$. As a result, this BH-boson cloud system acts like a continuous GW source~\cite{Arvanitaki:2014wva,Arvanitaki:2016qwi,Baryakhtar:2017ngi,Brito:2017wnc,Brito:2017zvb,Isi:2018pzk,Ghosh:2018gaw,Palomba:2019vxe,Sun:2019mqb,Zhu:2020tht,Brito:2020lup,Ng:2020jqd} and the incoherent superposition of the GWs from all the emitting sources in the Universe can lead to a strong stochastic GW background (SGWB)~\cite{Brito:2017wnc,Brito:2017zvb,Tsukada:2018mbp,Zhu:2020tht,Tsukada:2020lgt}.

So far, the null detection of a SGWB by LIGO, rules out scalar bosons with masses $\sim [1.3,3.8]\times10^{-13}$eV \cite{Tsukada:2018mbp} and vector bosons with masses $[0.8,6.0]\times10^{-13}$eV \cite{Tsukada:2020lgt}, under optimistic assumptions for the BH population. From this SGWB, it is expected that LIGO could observe $\sim[10^{-13},10^{-12}]$ eV bosons in the most optimistic scenarios \cite{Brito:2017wnc,Brito:2017zvb,Tsukada:2018mbp}. However, constraints on ultralight scalar fields only considered the SGWB produced by the fundamental dipolar mode with $m=1$~\cite{Tsukada:2018mbp}. This underestimates the total SGWB and hence could also affect the detectable mass range of bosons. The impact of the next most unstable mode with $m=2$ on the SGWB was explored in the case of ultralight vector bosons in~\cite{Tsukada:2020lgt}, but mostly focused on their impact for LIGO detection. Their work shows that with current sensitivities the impact of higher modes is limited, but suggests that it might become quite important for frequencies where future ground-based detectors might be orders of magnitude more sensitive than LIGO.

With this motivation in mind, we explore the impact of higher modes, focusing on a minimally coupled scalar field and studying in detail the impacts of generic higher modes with $m>1$ on the SGWB. We then explore, for the first time, the capability of future ground-based detectors, such as the Neutron Star Extreme Matter Observatory (NEMO) \cite{Ackley:2020atn}, Cosmic Explorer (CE) \cite{Evans:2016mbw} and the Einstein Telescope (ET) \cite{Punturo:2010zz,Maggiore:2019uih}, to detect this background. We argue that for bosons with masses $\gtrsim 10^{-12}$ eV, the SGWB is entirely dominated by the higher modes, which is particularly important for future detectors. Throughout this work we consider a cosmological $\Lambda$CDM model with cosmological parameters $\Omega_{m}=0.3$, $\Omega_{\Lambda}=0.7$ and $H_0=69.8 \mathrm{km /s/Mpc}$.

\section{Superradiant instability and GW emission}
We consider a real massive scalar field $\Phi$, minimally coupled to gravity, and neglect any possible coupling with other particles, as well as possible self-interactions.
We also neglect possible non-gravitational interactions of the bosonic field. Our results are robust as long as such non-gravitational interactions are sufficiently weak. For strong non-gravitational interactions we expect the final SGWB to be suppressed, but the specific details will depend on the model and requires further investigation~\cite{Yoshino:2015nsa,Rosa:2017ury,Ikeda:2018nhb,Boskovic:2018lkj,Fukuda:2019ewf,Mathur:2020aqv,Baryakhtar:2020gao,Omiya:2020vji}.
Take a BH spacetime, of mass $M$ and angular momentum $J=M^2\chi$. Linearized massive scalar fields can have quasi-bound state solutions of the form $\Phi=R(r)e^{-i\omega t+im\varphi}S_{lm}(\theta,\varphi)$. Here, we parametrized a two-dimensional unit sphere with angles $(\theta,\varphi)$, $S_{lm}$ is a spheroidal harmonic~\cite{Berti:2005gp} and $R(r)$ a radial wavefunction.
The eigenfrequencies $\omega$ belong to an infinite discrete set of complex numbers, namely (see e.g.~\cite{Brito:2015oca}):
\e
\omega_{nlm}\equiv \omega_R+i\omega_I\,,
\q
where $n$ is the principal quantum number, $l,m$ are quantum numbers related to the field angular momentum (and which describe the eigenvalues of the spheroidal harmonics) and $\omega_R,\omega_I$ are real quantities. An unstable state corresponds to $\omega_I>0$. Such modes exist whenever the frequency $\omega_R$ satisfies $0<\omega_R<m\Omega_{\mathrm{H}}$. Whenever the BH admits such modes, the BH transfers mass and spin to the boson field until the saturation point where $\omega_R\sim m\Omega_\mathrm{H}$. It takes a characteristic e-folding timescale, 
\begin{equation}
\tauinst\equiv \frac{1}{\omega_I}\,,
\end{equation}
for this saturation to occur and for the boson cloud to grow to a maximum mass, $\MSmax$. The instability timescale for modes with $l=m$ is always much shorter than those modes with $m<l$. Therefore we consider $l=m$ throughout and will mostly only refer to these modes by their azimuthal number $m$.

After this initial superradiant growth phase, the rotating boson cloud is dissipated through the emission of GWs with a typical half-life timescale $\taugw$ given by
\e
\tau_{\mathrm{GW}} = M_{f}\left(\frac{d \tilde{E}}{d t} \frac{M_{S}^{\max }}{M_{f}}\right)^{-1},
\q
where $M_f$ denotes the BH mass after saturation and $\frac{d \tilde{E}}{d t}\equiv \dot{E}_\text{GW} M_f^2/M_S^2$, with $\dot{E}_\text{GW}$ the GW flux which can be computed using techniques from BH perturbation theory~\cite{Yoshino:2013ofa,Brito:2017zvb,Siemonsen:2019ebd}\footnote{BH perturbation theory provides a good framework to compute GW emission because, in general, the curvature induced by the bosonic field remains small, compared to the BH curvature, during the whole evolution of the superradiant instability \cite{Brito:2014wla,East:2017ovw,Herdeiro:2017phl}.}. Since we are interested in generic $l=m$ modes we use the results obtained in~\cite{Yoshino:2013ofa} for massive scalar fields:
\e\label{Edot}
	\frac{d \tilde{E}}{d t}\approx\frac
	{
		16^{l+1} l(2 l-1) \Gamma(2 l-1)^{2} \Gamma(l+n+1)^{2}
		(\mu M)^{4l+10}
	}
	{
	n^{4 l+8}(l+1) \Gamma(l+1)^{4} \Gamma(4 l+3) \Gamma(n-l)^{2}
	},
\q
with $\Gamma$ denoting the Gamma function. We note that Eq.~(\ref{Edot}) is formally valid for $M\mu\ll l$, but provides a reasonably good approximation also for $M\mu \sim l$ \cite{Yoshino:2013ofa}.

Typically, $\taugw \gg \tauinst \gg M$, and therefore it is safe to assume that most of the GW emission takes place only after the boson cloud reaches the saturation point and that the system can be well described using a quasi-adiabatic approximation \cite{Brito:2014wla}. Let $M_i,\chi_i(M_f,\chi_f)$ denote the initial (final) mass and initial (final) dimensionless spin parameter of the BH, then we have $\MSmax\approx M_i-M_f$ and the final mass of the BH is given by \cite{Tsukada:2018mbp}
\e
M_{f}=\frac{m^{3}-\sqrt{m^{6}-16 m^{2} \omega_{R}^{2}\left(m M_{i}-\omega_{R} J_{i}\right)^{2}}}{8 \omega_{R}^{2}\left(m M_{i}-\omega_{R} J_{i}\right)}\,.
\q
Conservation of energy and angular momentum also yields
\e
J_f = J_i -{m\over \omega_R}(M_i-M_f).
\q
The total GW energy emitted by the boson cloud during a time $\Delta t$ can then be estimated using \cite{Brito:2017zvb}
\e\label{eq:energy}
E_{\mathrm{GW}}=\frac{M_{S}^{\max } \Delta t}{\Delta t+\tau_{\mathrm{GW}}}.
\q
Since we are interested in the GWs observed today, then we have $\Delta t=t_0-t_f$ representing the signal duration time~\cite{Tsukada:2020lgt}. Here $t_0$ is the age of the Universe today and $t_f$ is the cosmic time at the cosmological redshift where the BH forms.

\subsection{Contribution of modes $m>1$}
%
\begin{figure*}[htbp!]
	\centering
	\includegraphics[width = 0.49\textwidth]{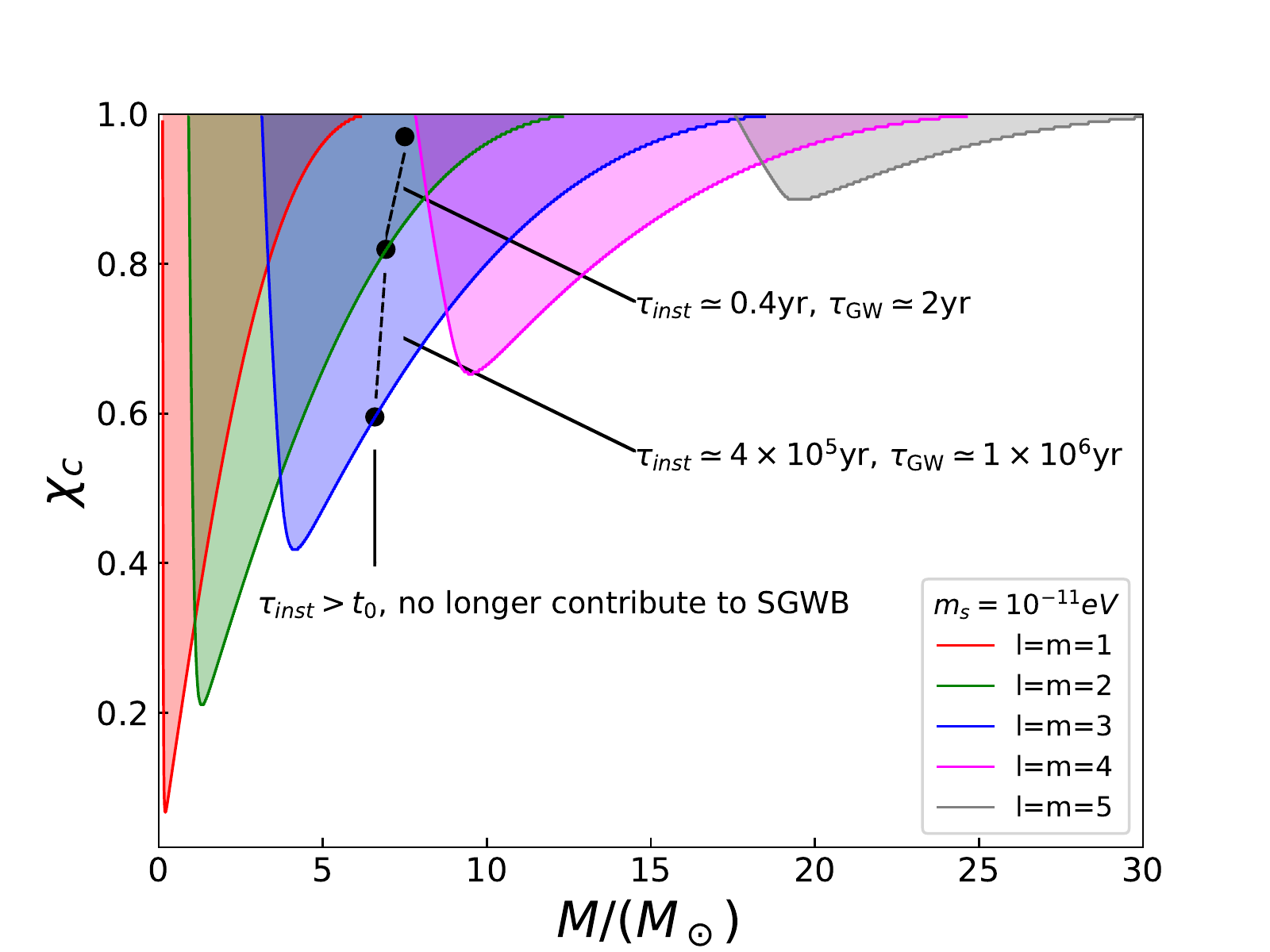}
	\includegraphics[width = 0.49\textwidth]{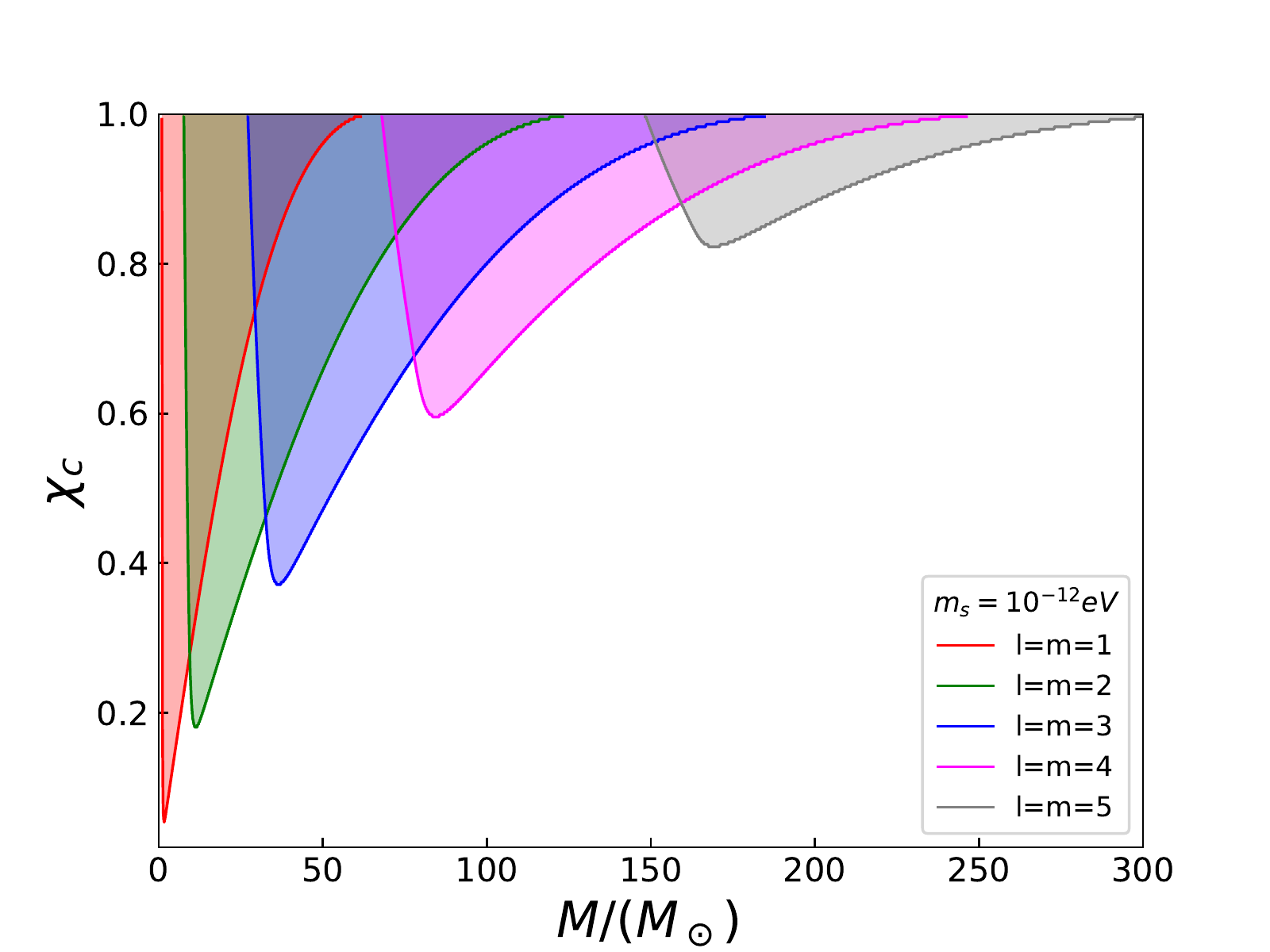}
	\caption{\label{critspin} The critical spin $\chi_c$ (colored lines) of BHs with mass $M$ above which a superradiant instability for scalar fields with mass $m_s=10^{-11}$ eV (left panel) and $m_s=10^{-12}$ eV (right panel) can be triggered and for which the corresponding instability timescale is shorter than the age of the Universe, for several values of $l=m$. A BH born inside the colored regions would be unstable on a timescale much shorter than age of Universe and lose spin until reaching the critical spin. For illustration, in the left panel, we show an example trajectory for a BH born inside an unstable region.
	}
\end{figure*}
So far, our discussion assumes that only a single mode becomes unstable over the BH lifetime. However, in general, any modes with instability timescale shorther than the BH age can, in principle, follow the exact same process outlined above and contribute to the overall GW emission.
Assume, for simplicity, that $\mu M\ll 1$. In this limit the growth rate of the superradiant instability scales as $\omega_I\propto {(\mu M)}^{4l+4}$~\cite{Detweiler:1980uk}. Since $\mu M\ll 1$, the instability timescale grows quickly with increasing $l$, and therefore it is safe to assume that each mode grows one at a time (see e.g.~\cite{Ficarra:2018rfu}). This hierarchy of timescales between different modes can be shown to hold also beyond the $\mu M\ll 1$ limit~\cite{Dolan:2007mj,Eperon:2019viw}. Therefore, to compute the contribution from different modes to the total GW emission we use the following procedure:
\begin{itemize}
	\item For a given initial BH mass and spin, compute the final mass $M_{k}$ and spin $\chi_{k}$, after the evolution of modes $m=1,2,\cdots,k$;
	\item The total GW energy from the first $k$ modes is $E_\mathrm{GW}=\sum_{j=1}^{k}E_\mathrm{GW}(l=m=j)$, with $E_\mathrm{GW}(l=m)$ computed using Eq.~\eqref{eq:energy};
	\item Check if the next mode $m=k+1$ can trigger a superradiant instability for a BH with mass $M_{k}$ and spin $\chi_{k}$ on a timescale shorter than $\Delta t$. If not, then the total GW energy is just $E_\mathrm{GW}$ computed in the previous step. Otherwise, the BH will continue evolving until reaching mass $M_{k+1}$ and spin $\chi_{k+1}$. The total GW energy becomes $E_\mathrm{GW}=\sum_{j=1}^{k+1}E_\mathrm{GW}(l=m=j)$.
	\item Start from $m=1$ and repeat the above steps until a given mode does not contribute. End the computation when this happens. The total emitted GW energy is then given by $E_\mathrm{GW}$ as computed in the last step of this procedure.
\end{itemize}

For instance, we first evaluate the GWs generated by the most unstable mode $m=1$ in the absence of all higher modes. When the mode $m=1$ reaches saturation, the mass and spin of the BH evolve from $M_i$ and $\chi_i$ to $M_{m=1}$ and $\chi_{m=1}$. Then, we evaluate if the superradiant instability of a $M_{m=1}$ and $\chi_{m=1}$ BH can be triggered by the next most unstable mode $m=2$ and check if the instability timescale of this mode is shorter than the $\Delta t$. If yes, we evaluate the GW emission from this mode and add it to total energy budget. We continue this process until we find a mode for which the instability timescale is larger than $\Delta t$. 
For this computation we use the analytical expressions for $\omega_R$ and $\omega_I$ obtained in \cite{Eperon:2019viw}. These analytical approximations were obtained within a $l \gg 1$ limit, but provide an extremely good estimate to full numerical results also for small $l$ and for $M\mu \sim l$ \cite{Eperon:2019viw}. 

The above procedure can be better understood by looking at Fig.~\ref{critspin}, where we show the regions where different modes are unstable on timescales shorter than $\Delta t$ and therefore can contribute to the overall GW emission. For illustration, we show the specific cases of scalar fields with mass $m_s=10^{-11}$ eV (left panel) and  $m_s=10^{-12}$ eV (right panel). A BH formed inside the colored regions would be unstable against the corresponding $m$-mode and evolve until reaching the critical spin where the instability stops for the corresponding unstable mode, leading to the creation and subsequent GW emission from a boson cloud with quantum numbers $m$. Figure~\ref{critspin} also shows that the modes relevant for GW emission depend on both the BH initial mass and spin but also on the scalar field's mass. For example, one can see that for $m_s=10^{-11}$ eV and BHs born with mass $M_i\sim  5\, M_{\odot}$ and spin $\chi_i=0.7$, only modes with $m>1$ can become superradiantly unstable, whereas such BHs would still be superradiantly unstable against the $m=1$ mode for $m_s=10^{-12}$ eV.

As a final comment, we note that in our computation we only consider modes with principal quantum number $n=l+1$, corresponding to the fundamental tone of each mode. This is because for larger $n$-modes the critical spin above which they are unstable is always larger than the corresponding critical spin for the fundamental mode. Therefore even though some overtones might grow on timescales comparable to the fundamental mode, their contribution to the total GW emission is negligible, since they are quickly re-absorbed by the BH when the fundamental mode reaches saturation~\cite{Siemonsen:2019ebd}. For simplicity, in the following part, we will write $m$-mode to represent the $(n,l,m)=(m+1,m,m)$ mode.

\section{Stochastic GW background}

The incoherent superposition of unresolvable GWs from a population of BHs will form a SGWB. For boson clouds this has been studied in Refs.~\cite{Brito:2017wnc,Tsukada:2018mbp,Tsukada:2020lgt} for extragalactic sources and also in~\cite{Zhu:2020tht} for galactic sources. These works mainly focused on the most unstable mode $m=1$, with the exception of~\cite{Tsukada:2020lgt} where $m=2$ modes were also taken into account, and only considered the detectability of this SGWB with LIGO. 
In this Section, we compute the SGWB from extra-galactic sources, considering the contribution from any modes $m\geq 1$ as described above. We consider both stellar-origin BHs (SBHs) and, more hypothetical, primordial BHs (PBHs). Unlike SBHs, PBHs could also allow the existence of BHs with masses with $M\lesssim 3\, M_{\odot}$, and therefore we consider such a population to understand whether, given current constraints on their existence, the existence of PBHs could allow to detect a SGWB from boson clouds from larger boson masses compared to SBHs.

We take into account both BHs formed in isolation from stellar collapse or formed in binary BH mergers. Since PBHs are formed from the critical collapse of all the matter inside a Hubble volume at horizon entry \cite{Hawking:1971ei,Carr:1974nx} they are generically expected to be formed with negligible spin \cite{Chiba:2017rvs,DeLuca:2019buf}, and even when considering gas accretion, the spin of PBHs is still expected to be negligible if $M\lesssim 30\Msun$~\cite{DeLuca:2020bjf}. Therefore we do not consider isolated PBHs, since for most isolated PBHs superradiant instabilities are unlikely to be triggered. Although some PBH models also predict fast spinning PBHs (see e.g., \cite{Cotner:2019ykd}), we only consider the PBH remnants from two non-spinning PBHs in order to obtain the most conservative estimate on the detectablity of such a background. Hence, for PBHs, we only consider PBH remnants which are formed in binary PBH mergers.

Finally, we should note that we do not consider the contribution from isolated BHs in the galaxy~\cite{Zhu:2020tht}, since this component is expected to add a non-isotropically distributed component to the SGWB, which would require a different treatment. In addition, the GW signals emitted by a galactic BH population are expected to accumulate in a narrow frequency window around $f_0\sim \mu/\pi$, unlike the extra-galactic component which, as we show below, is spread over a broader range of frequencies due to the cosmological redshift. The SGWB search methods that we focus on, assume a mostly Gaussian and isotropically distributed signal that emit in a broad range of frequencies, such as the one that is expected to come from the extragalactic population~\cite{Tsukada:2020lgt}. 

\subsection{Background from isolated BHs}

The SGWB can be characterized in terms of the dimensionless energy density parameter (see e.g.~\cite{Maggiore:1900zz})
\e
\Omega_{\mathrm{GW}}(f)\equiv {1\over \rho_c}\frac{d\rho_{\mathrm{GW}}}{d \ln(f)},
\q
defined as the GW energy density per logarithm frequency, normalized by the critical energy, $\rho_c$, needed for closing the Universe. This quantity can be computed by summing over all the GW sources present in the sky which, for sources isotropically distributed in the sky, leads to \cite{Phinney:2001di}:
\e
\Omega_{\mathrm{GW}}(f)=\frac{f}{\rho_{c}} \int d \chi d M d z \frac{d t_L}{d z} \frac{d^{2} \dot{n}}{d M d \chi} \frac{d E_{s}}{d f_{s}}\,.
\q
Here $d t_L/dz$ stands for the derivative of the lookback time with respect to the cosmological redshift and ${d^{2} \dot{n}}/{d M d \chi}$ denotes the BH formation rate per comoving volume per BH mass per spin in the source frame.

Since our GW sources emit at nearly constant frequencies, the energy spectrum of a single GW event in the source frame, $d E_s/d f_s$, is very well approximated by a delta function~\cite{Brito:2017wnc,Tsukada:2018mbp,Tsukada:2020lgt},
\e
\frac{d E_{s}}{d f_{s}} = E_{\mathrm{GW}} \delta\left(f(1+z)-f_{0}\right).
\q
For SBHs, we compute the BH formation rate as a function of redshift using~\cite{Brito:2017wnc,Tsukada:2018mbp,Tsukada:2020lgt}
\e\label{ndot}
\frac{d \dot{n}}{d M}= \int \psi[t-\tau(M_*)]\phi(M_*)\delta(M_*-g^{-1}_{\mathrm{rem}}(M,z))\mathrm{d}M_*,
\q
where $M_*$ is the mass of the progenitor star and its lifetime is given by $\tau(M_*)$, which can be found in~\cite{Schaerer:2001jc}. We adopt the Salpeter initial mass function $\phi(M_*)\propto M_*^{-2.35}$ \cite{Salpeter:1955it} and normalize it in the mass range $[0.1,100]\Msun$. For a progenitor with mass $M_*$, the mass of the BH remnant $M$ is given by $M=g_{\mathrm{rem}}(M_*,z)$ and we use the approach in \cite{Fryer:2011cx} to evaluate the function $g_{\mathrm{rem}}$. We set the solar metallicity in the expression of $g_{\mathrm{rem}}$ to be $Z_\odot=0.0196$ \cite{Vagnozzi:2017wge} and the metallicity as a function of $z$ is taken from \cite{Belczynski:2016obo}. For the star formation rate, $\psi(z)$, we use the functional form \cite{Springel:2002ux}
\e
\psi(z)=\nu\frac{a\exp[b(z-z_m)]}{a-b+b\exp[a(z-z_m)]},
\q
and we adopt the fitted parameters given in \cite{Vangioni:2014axa}, $\nu=0.178\Msun \mathrm{yr}^{-1}\mathrm{Mpc}^{-3}$, $z_m=2.00$, $a=2.37$ and $b=1.80$\footnote{As shown in~\cite{Tsukada:2020lgt}, different choices of the star formation rate, given observational uncertainties, do not greatly affect the overall amplitude of the SGWB.}. Moreover, we set SBHs formed through stellar collapse to have masses within $M\in[3,50]\Msun$. Finally, since the above model for SBHs does not include spin, we follow \cite{Brito:2017zvb,Brito:2017wnc,Tsukada:2018mbp,Tsukada:2020lgt} and model the spin with a uniform distribution in the range $\chi_i\in[0,1]$. To gauge the uncertainty in the amplitude of the SGWB,  we also consider the most extreme scenario where all isolated BHs are born with negligible spin in which case only BHs formed from binary BH mergers contribute to the background.

\subsection{Background from BH remnants}
Let us now consider the SGWB from BH remnants formed from binary BH mergers. Unlike the case of isolated BHs for which the spin distribution at birth is highly uncertain, the spin distribution of BH remnants is expected to be, on average, peaked around $\chi\approx 0.7$~\cite{Gerosa:2017kvu,Fishbach:2017dwv}, due to the fact that the merger of nearly equal-mass, slowly-spinning BHs tend to form  BHs with such a dimensionless spin~\cite{Berti:2007fi,Scheel:2008rj,Barausse:2009uz}. Accurate expressions from Numerical Relativity simulations predict that the mass $M_i$ and dimensionless spin $\chi_i$ of a BH arising from the merger of two non-spinning BHs of masses $M_1,M_2$ are~\cite{Berti:2007fi,Scheel:2008rj,Barausse:2009uz}
\begin{eqnarray}\label{massspin}
M_i&=&M_1+M_2-(M_1+M_2)\Bigg[\left(1-\sqrt{\frac{8}{9}} \right)\nu\nonumber\\
&\qquad&-4\nu^2\left(0.19308+\sqrt{\frac{8}{9}} -1 \right)\Bigg],\nonumber\\
\chi_i&=&\nu (2\sqrt{3}-3.5171\nu+2.5763\nu^2)\,,
\end{eqnarray}
with $\nu \equiv M_1M_2/(M_1+M_2)^2$ the symmetric mass ratio.

The energy density parameter can then be evaluated using \cite{Phinney:2001di}
\e
\Omega_{\mathrm{GW}}(f)=\frac{f}{\rho_{c}} \int  d M_1 dM_2 d z \frac{d t}{d z} \mathcal{R}(z,M_1,M_2) \frac{d E_{s}}{d f_{s}},
\q
where $\mathcal{R}(z,M_1,M_2)$ is the merger rate density for binary BHs with component mass $M_1$ and $M_2$ (we require $M_1\ge M_2$) and, following our assumptions, $d E_{s}/d f_{s}$ is computed using Eq.~(\ref{massspin}). We neglect the impact of the spin of the BH progenitors, since LIGO and Virgo observations seem to indicate that they tend to be, typically, small~\cite{Abbott:2020gyp}.

Let us first consider BH remnants formed due to the merger of SBHs. We follow \cite{Abbott:2017xzg,TheLIGOScientific:2016wyq} to obtain the merger rate density of binary SBHs. The merger rate density $\mathcal{R}(t,M_1,M_2)$, is given by the convolution of the SBH formation rate and the time delay $t_d$ between formation and merger:
\e\label{eq:rateSBH}
\mathcal{R}\propto \int_{t_{\min }}^{t_{\max }}
\frac{d \dot{n}}{d M}
\left(t-t_{d}, M_1\right) P_{d}\left(t_{d}\right)P(M_1)P(M_2) d t_{d}\,.
\q
Here $P(M_1)$ and $P(M_2)$ are the mass functions of the component masses, and $P_{d}\left(t_{d}\right)$ is the time delay distribution. 
Similarly to \cite{Abbott:2017xzg,TheLIGOScientific:2016wyq}, we adopt $t_{\min} =50$Myr, $t_{\max}$ to be the Hubble time and consider a time delay distribution given by $P_d(t_d)\propto t_d^{-1}$. 
Following the constraints imposed by LIGO and Virgo's latest observations we fix the local merger rate, $\int \mathcal{R}(z=0,M_1,M_2)dM_1dM_2$, to $23.9\, \mathrm{yr}^{-1}\mathrm{Gpc}^{-3}$. For the mass functions, we adopt the ``broken power law mass model'' in \cite{Abbott:2020gyp} and use the median values in their Fig.~17 for the free parameters.

Finally, for the more hypothetical PBH population, we consider the merger rate density computed in~\cite{Chen:2018czv}, valid for an arbitrary PBH mass function:
\m\label{eq:ratePBH}
	\mathcal{R}&\approx & 3.9 \cdot 10^{6}\mathrm{Gpc}^{-3}\,\mathrm{yr}^{-1} \times\left(\frac{t}{t_{0}}\right)^{-\frac{34}{37}} f^{2}\left(f^{2}+\sigma_{\mathrm{eq}}^{2}\right)^{-\frac{21}{74}} \no\\
	&\times & \min \left(\frac{P\left(M_{1}\right)}{M_{1}}, \frac{P\left(M_{2}\right)}{M_{2}}\right)\left(\frac{P\left(M_{1} \right)}{M_{1}}+\frac{P\left(M_{2} \right)}{M_{2}}\right) \no\\
	&\times &\left(M_{1} M_{2}\right)^{\frac{3}{37}}\left(M_{1}+M_{2}\right)^{\frac{36}{37}},
\n
where $P(M_i)$, $i=1,2$, is the PBH mass function, normalized such that $\int P(M_i)d M_i=1$, and $\sigma_{\mathrm{eq}}\approx0.005$ \cite{Ali-Haimoud:2017rtz}. 
The parameter $f\equiv \Omega_{\mathrm{PBH}}$ is related to the fraction of PBH DM by $\fpbh=f/\Omega_{\mathrm{DM}}\approx f/0.85$, where $\Omega_{\mathrm{DM}}$ ($\Omega_{\mathrm{PBH}}$) is the fraction of the total energy density of our Universe that is composed of DM (PBHs). 

Various independent constraints on $\fpbh$, indicate that the current maximum allowed $\fpbh$ in the solar mass window -- which is the most relevant for our purposes -- is no more than a few part in thousand \cite{Bird:2016dcv,Chen:2019irf,Carr:2020gox}. We are also mostly interested in seeing the impact on the SGWB that a possible population of PBHs with masses below the ones that can be formed through stellar collapse would have. Therefore, for concreteness, we only consider PBHs with masses $\lesssim 5\, M_{\odot}$. On the other hand, recent results suggest that the existence of PBHs with mass $ \sim [10^{-3}, 1]\, M_{\odot}$ is strongly constrained by pulsar timing array data~\cite{Chen:2019xse}. Therefore, we will take the maximum allowed abundance,  $\fpbh=10^{-3}$ in the mass range $[1,5]\,\Msun$. See e.g., Fig.~7 in \cite{Yuan:2021qgz} for current constraints on $\fpbh$. We adopt a power-law mass function $P(M_1)=P(M_2)\propto M^{-3/2}$, where the power-law index $-3/2$ corresponding to PBHs generated by a broad and flat primordial scalar power spectrum \cite{DeLuca:2020ioi}.  We also consider a log-normal mass function, namely
\begin{equation}\label{log}
P(M)=\frac{1}{\sqrt{2 \pi} \sigma M} \exp \left(-\frac{\ln ^{2}\left(M / M_{p}\right)}{2 \sigma^{2}}\right),
\end{equation}
where $M_p$ denotes the peak mass and $\sigma$ is the width of the mass spectrum. We set $M_p=3\Msun$ and $\sigma=0.5$, such that the relevant BH mass range is roughly the same as for the power-law mass function.

\section{Observational prospects}
%
\begin{figure*}[htbp!]
	\centering
	\includegraphics[width = 0.49\textwidth]{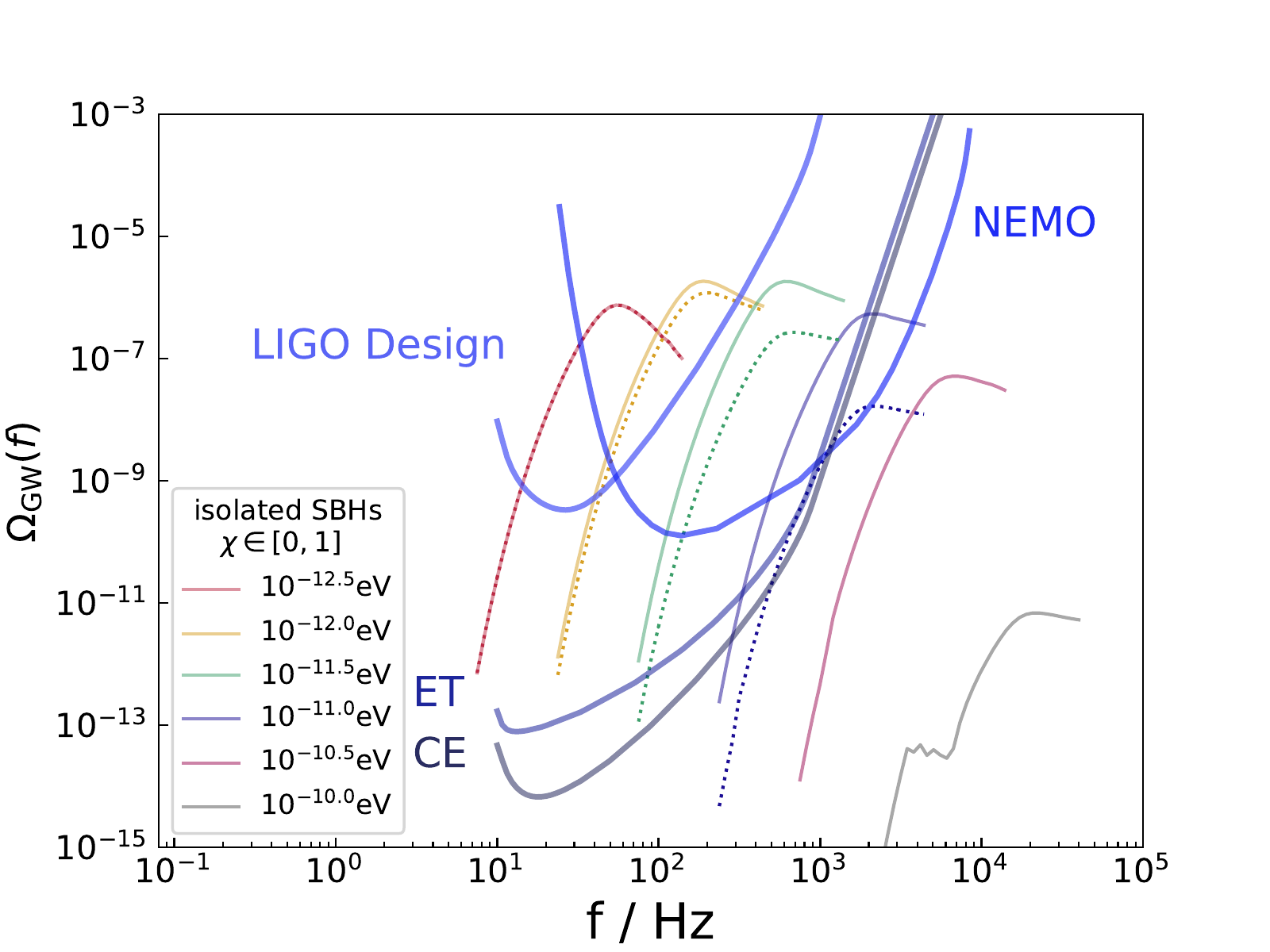}
	\includegraphics[width = 0.49\textwidth]{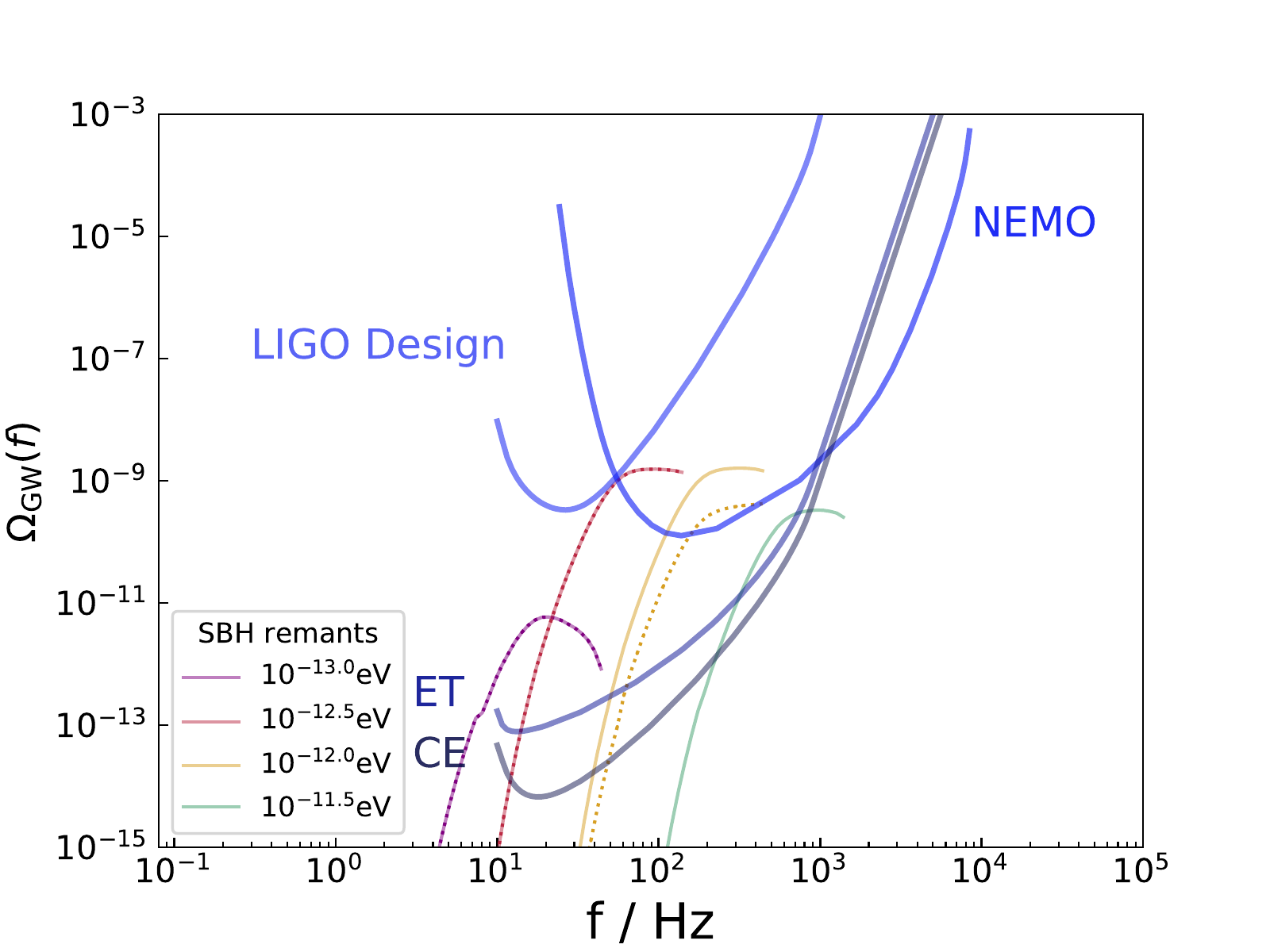}
	\caption{\label{highermodes} Left panel: Stochastic GW background from scalar clouds formed around SBHs assuming a uniform initial spin  distribution $\chi_i \in [0,1]$ for BHs formed in isolation. Right panel: Same but only taking into account the contribution from BHs formed through binary SBH mergers.
For both panels, the dotted lines only include the mode $m=1$ while the solid lines include $m\geq 1$ with all the $m$-modes that contribute. We also plot the power-law integrated sensitivity curves~\cite{Thrane:2013oya} by assuming the threshold SNR to be $\SNR=1$.
For LIGO/NEMO/CE/ET at their design sensitivity. We assume a four-year-detection and two co-aligned and co-located identical detectors for NEMO/CE/ET, whereas for LIGO we use the overlap function computed in~\cite{Thrane:2013oya}. 
}
\end{figure*}

We are now in position to study the prospects to observe a SGWB with current and future ground-based GW detectors. Let us first focus on the SGWB emitted due to the superradiant instability of SBHs, either formed in isolation due to stellar collapse or formed through binary BH mergers.
In the left panel of Fig.~\ref{highermodes} we show the SGWB when summing over the two formation channels, computed following the procedure outlined in the previous Section. As mentioned above, we recall that we assume a uniform BH spin distribution $\chi_i \in [0,1]$ for the isolated BH channel. To highlight the importance of adding modes $m>1$ in the computation we show the results both when considering only the $m=1$ mode (dotted lines) and when considering all modes $m\geq 1$ (solid lines) for which the instability timescale is much shorter than $\Delta t$. As one can see, higher $m$-modes are particularly important for heavier bosons, especially for $m_s \gtrsim 10^{-12}$ eV, in agreement with~\cite{Tsukada:2020lgt}. We also note that for $m_s\lesssim 10^{-12.5}$ eV the importance of $m>1$ modes is negligible (hence for those cases the solid and dotted lines are indistinguishable in Fig.~\ref{highermodes}), whereas for $m_s\gtrsim 10^{-11}$ eV the background is entirely due to the $m>1$ modes.
These results are especially relevant for future GW detectors, since as one can infer from the left panel of Fig.~\ref{highermodes} and as we confirm below, the SGWB produced by such bosons would be only detectable with future detectors such as NEMO, ET or CE.

The relative importance of different modes can be understood from Fig.~\ref{critspin}, bearing in mind that the BHs in our SBH population have masses $\geq 3\,M_{\odot}$. Focusing, for example, on the SGWB for $m_s\sim 10^{-11}$ eV, it is clear from Fig.~\ref{critspin} that, given the constraint $\geq 3\,M_{\odot}$ for most BHs in the population only modes with $m>1$ can become superradiantly unstable and therefore contribute to the overall SGWB. For larger boson masses, the importance of higher modes becomes even more extreme, since for $m_s\gtrsim 10^{-11}$ eV no BHs in the population are unstable against the $m=1$ mode and therefore only higher modes contribute. For smaller scalar field masses instead, since most BHs in the population are unstable against the $m=1$ mode, and because this mode has the smallest instability and GW emission timescales, the importance of higher order modes becomes less important. We also note that, for all boson masses of interest, the SGWB depends at most on modes $m<9$ since the instability timescale for $m\geq 9$ is larger than $\Delta t$ for all BHs in the population. 

The results discussed so far assumed that BHs in isolation are born with sufficiently high spins such that they can be superradiantly unstable and produce a very large background. In fact, when assuming that isolated BHs are born with uniformly distributed between $\chi_i \in [0,1]$, the SGWB in Fig.~\ref{highermodes} is completely dominated by the isolated BH channel, in agreement with previous results~\cite{Tsukada:2018mbp,Tsukada:2020lgt}. Within such assumptions, the existence of bosons with masses close to $m_s \sim 3 \times  10^{-12}$ eV is actually already in tension with the fact that no SGWB as been detected on LIGO data so far. Let us therefore consider a worst-case scenario, in which all BHs formed through stellar collapse are born with zero or negligible spin. In such scenario, only BHs formed through binary BH mergers can become superradiantly unstable and contribute to the SGWB. In the right panel of Fig.~\ref{highermodes} we show the SGWB that would be produced in this case. As one can see, the amplitude of the background would be substantially smaller and the prospects for detection would be substantially better with future detectors.

\begin{figure*}[htbp!]
	\centering
	\includegraphics[width = 0.49\textwidth]{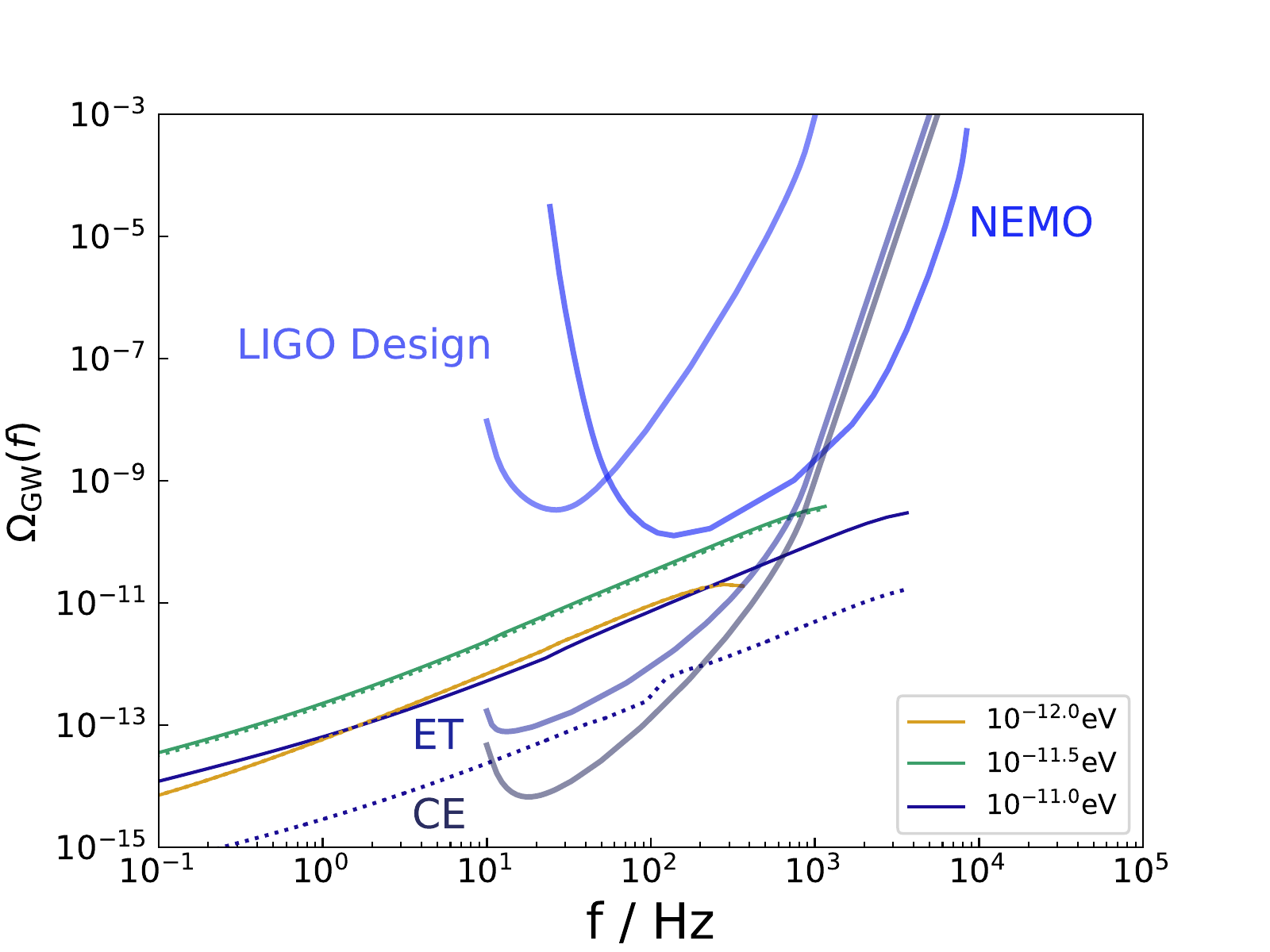}
	\includegraphics[width = 0.49\textwidth]{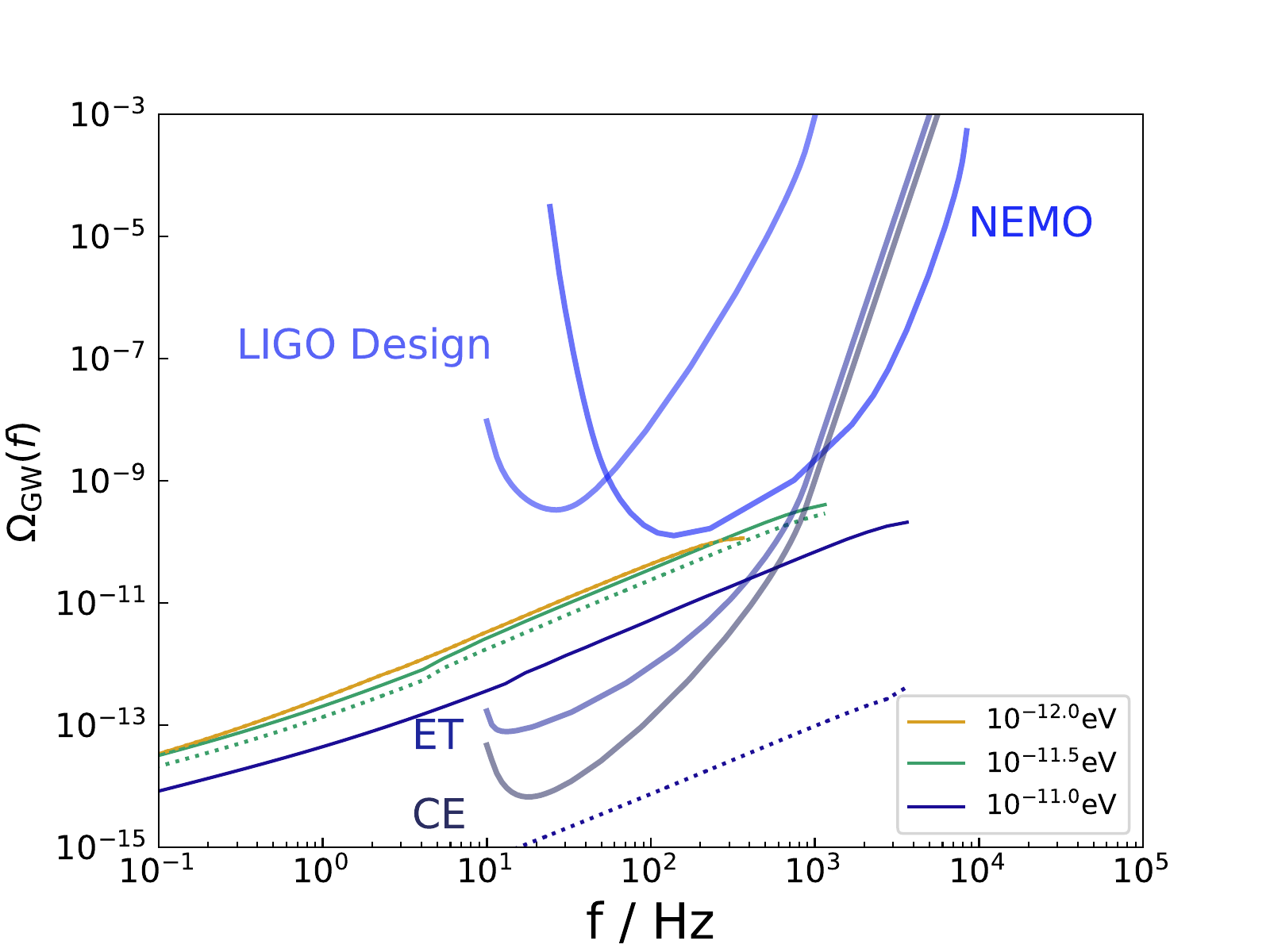}
	\caption{\label{ogwpbh} Same as Fig.~\ref{highermodes} but for the stochastic GW background from scalar clouds formed around PBH remnants. Upper panel: The mass function of PBHs is $P(M)\propto M^{-3/2}$ in the mass window $M\in [1,5]\Msun$.
	Lower panel: The mass function of PBHs has a log-normal form described by Eq.~(\ref{log}). We set $M_p=3\Msun$ and $\sigma=0.5$.
}
\end{figure*}

Finally, let us consider the case of PBH remnants, using the population model discussed in the previous Section. The SGWB spectrum for this case is shown in Fig.~\ref{ogwpbh}. Similarly to the case of SBHs (cf. Fig.~\ref{highermodes}), the background is also dominated by modes $m>1$ for heavier bosons, and for the same reasons outlined above. Some comments are in order here.  Because we only consider BHs in the mass window $M\in [1,5]\Msun$ for the power-law model, the SGWB is strongly suppressed for $m_s\lesssim 10^{-12} $ eV because for such bosons the superradiant instability timescale becomes much larger than $\Delta t$ for all BHs in the population. The same happens for the log-normal distribution for our choice of parameters $M_p=3\Msun$ and $\sigma=0.5$. Since PBHs are peaked around $M_p$, the SGWB is strongly suppressed for $m_s\lesssim10^{12.5}$ eV. In addition, given our assumptions for the BH population, this contribution to the background would be negligible in the frequencies at which LIGO is most sensitive. On the other hand, for $m_s\gtrsim 10^{-11}$ eV the contribution to the SGWB from such a PBH population can become more important than the background of SBHs due to the existence of BHs with masses $\lesssim 3\, M_{\odot}$. We also note that the different spectral index between the SGWB in Fig.~\ref{ogwpbh} compared to right panel of Fig.~\ref{highermodes} is mainly related to the different dependence of the merger rate on the cosmological redshift
(Eq.~\eqref{eq:rateSBH} and Eq.~\eqref{eq:ratePBH}).

\subsection{Signal-to-noise ratio}
To further quantify the prospects for detection and the impact of higher $m$-modes, we estimate the signal-to-noise ratio (SNR) with which this background could be observed. The SNR, valid for an arbitrary large SGWB, can be computed using~\cite{Allen:1997ad}:
\e\label{SNR}
\rho^{2}\!=T \! \int \! \mathrm{d} f \frac{\Gamma(f)^{2} S_{h}(f)^{2}}{\left[{1\over 25}+\Gamma(f)^{2}\right] S_{h}(f)^{2}+P_{n}(f)^{2}+{2\over 5} S_{h}(f) P_{n}(f)},
\q
where $\Gamma(f)$ and $P_n(f)$ are the overlap function and the noise power spectral density of the detector, respectively. The strain power spectral density of the SGWB, $S_h(f)$, is related to $\Omega_{\mathrm{GW}}(f)$ by $S_h(f) = 3H_0^2\Omega_{\mathrm{GW}}(f)/(2\pi^2 f^3)$. We assume an observation time $T=4$ yr.
Eq.~(\ref{SNR}) assumes two identical interferometers with a $90$ degree opening angle. Except for LIGO, we assume two co-aligned, co-located detectors for NEMO/CE/ET, where $\Gamma(0)=1/5$. In the weak signal limit, Eq.~\eqref{SNR} returns to the more commonly used expression \cite{Allen:1997ad,Thrane:2013oya}
\e
\rho^{2}\!\approx T \! \int \! \mathrm{d} f \frac{S_{h}(f)^{2}}{P_{n}(f)^{2}/R(f)^{2}}.
\q

\begin{figure*}[htbp!]
	\centering
	\includegraphics[width = 0.49\textwidth]{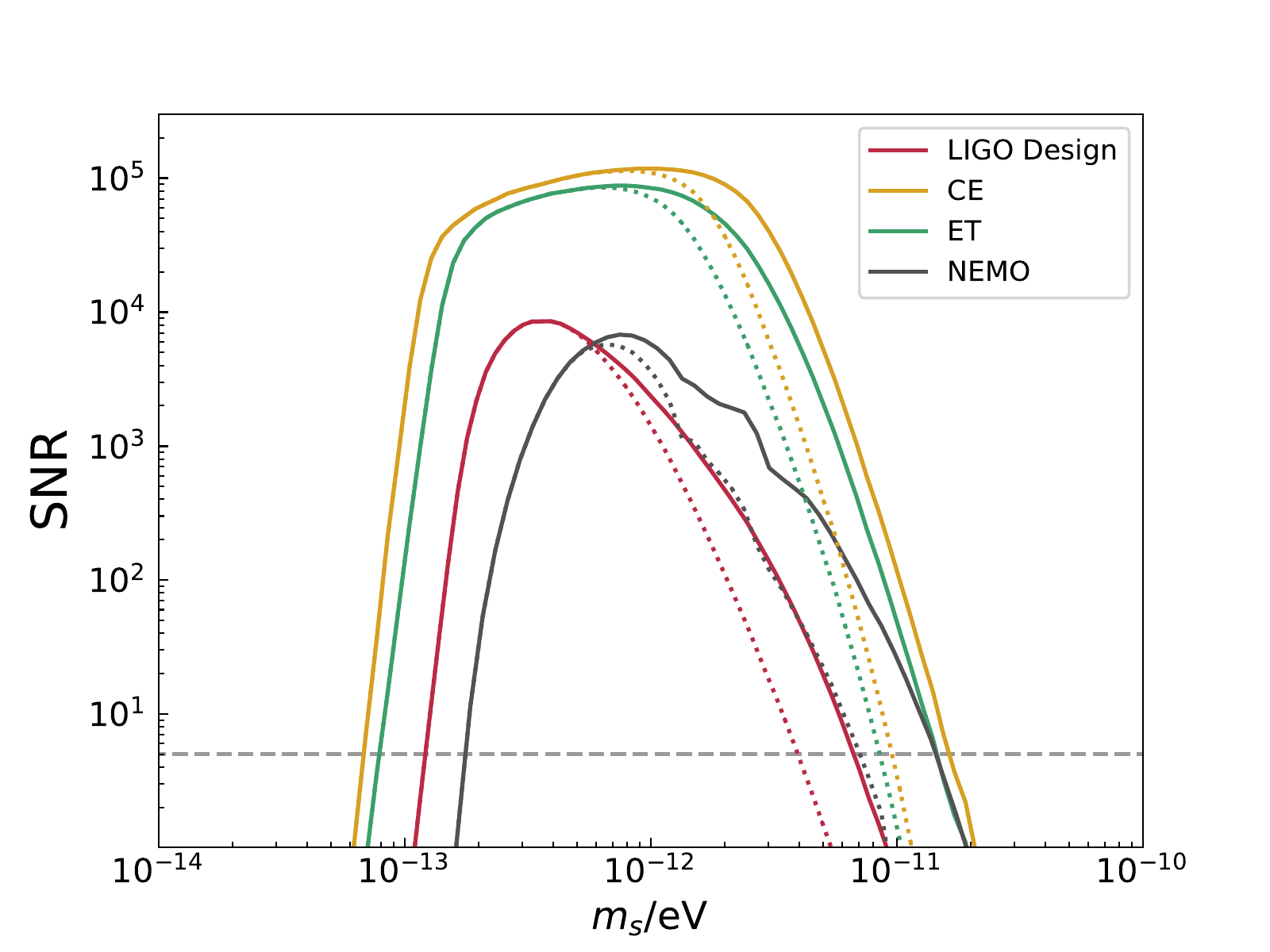}
	\includegraphics[width = 0.49\textwidth]{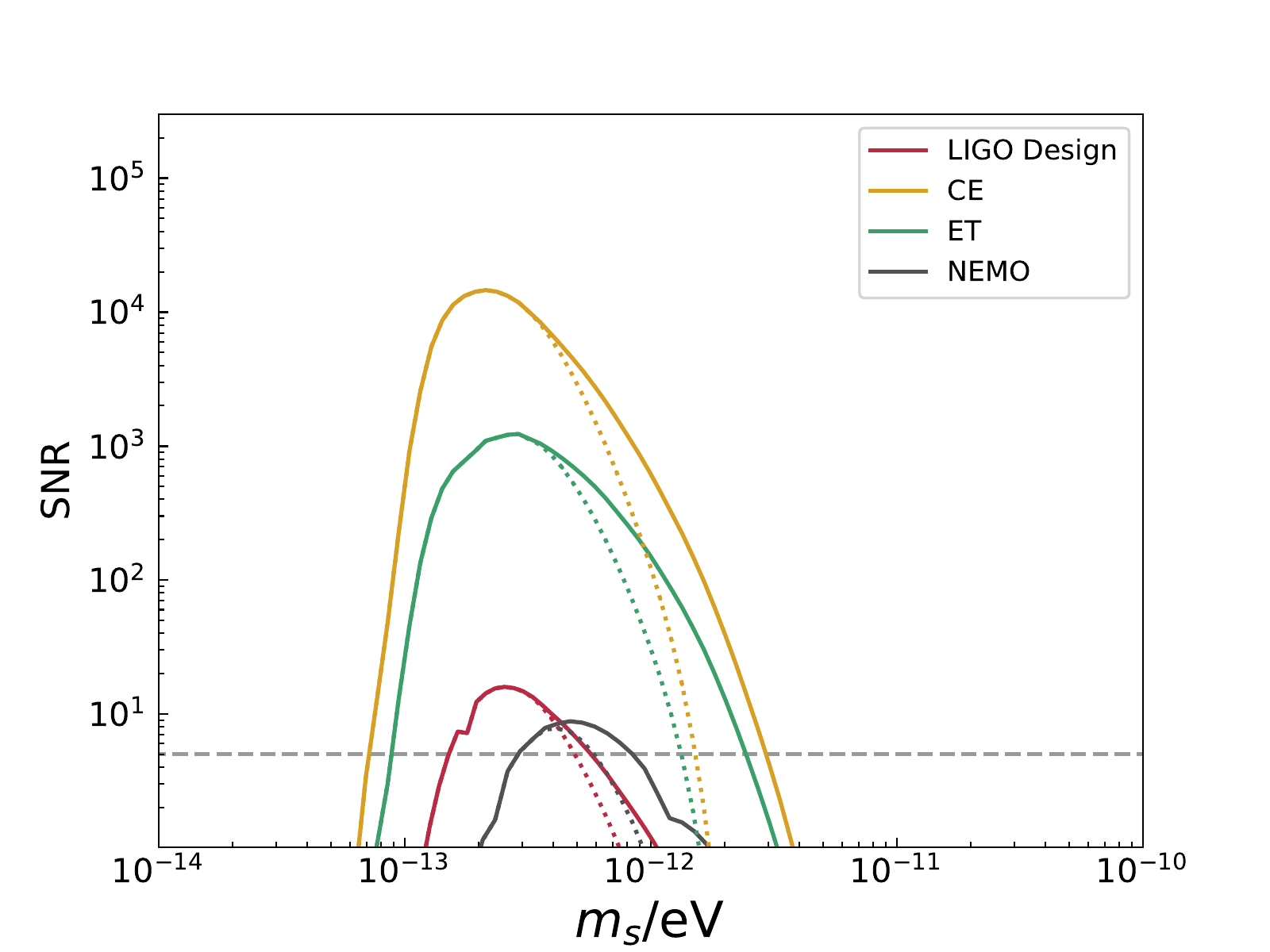}
	\caption{\label{SNRplot} Left panel: SNR for the SGWB, as a function of the boson mass $m_s$, from stellar-origin BHs taking into account both isolated and remnant BHs and assuming a uniform initial spin distribution $\chi_i \in [0,1]$ for BHs formed in isolation. Right panel: SNR for the SGWB when only taking into account the contribution for BHs formed through the merger of stellar-origin BHs.
The dashed line corresponds to $\mathrm{SNR}=5$. For all the detectors, we assume an observation time $T=4$ yr.
Dotted lines only include the fundamental mode $m=1$ while solid lines include $m\geq 1$ with all the $m$ that contribute.
	}
\end{figure*}
\begin{figure*}[htbp!]
	\includegraphics[width = 0.49\textwidth]{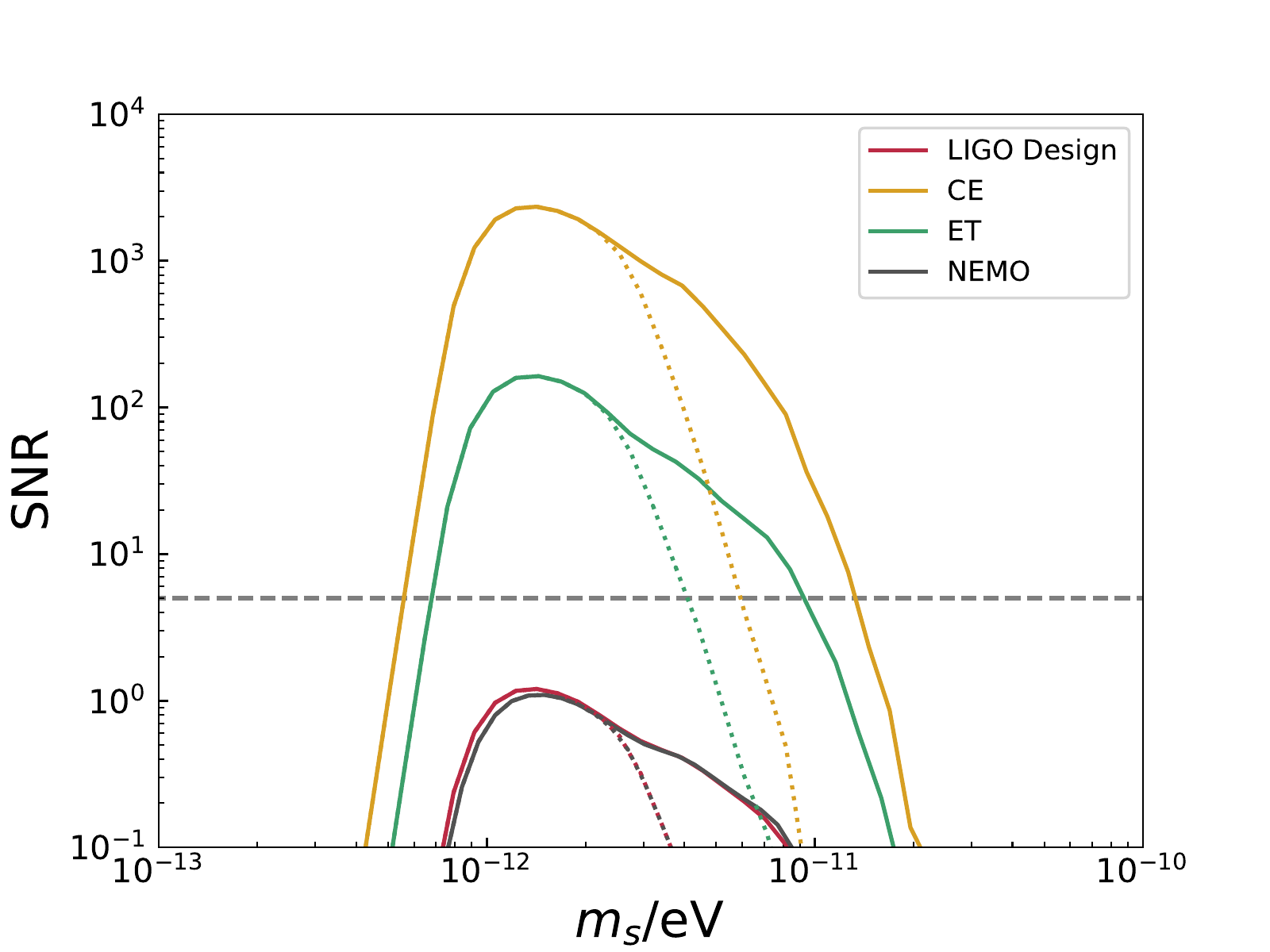}
	\includegraphics[width = 0.49\textwidth]{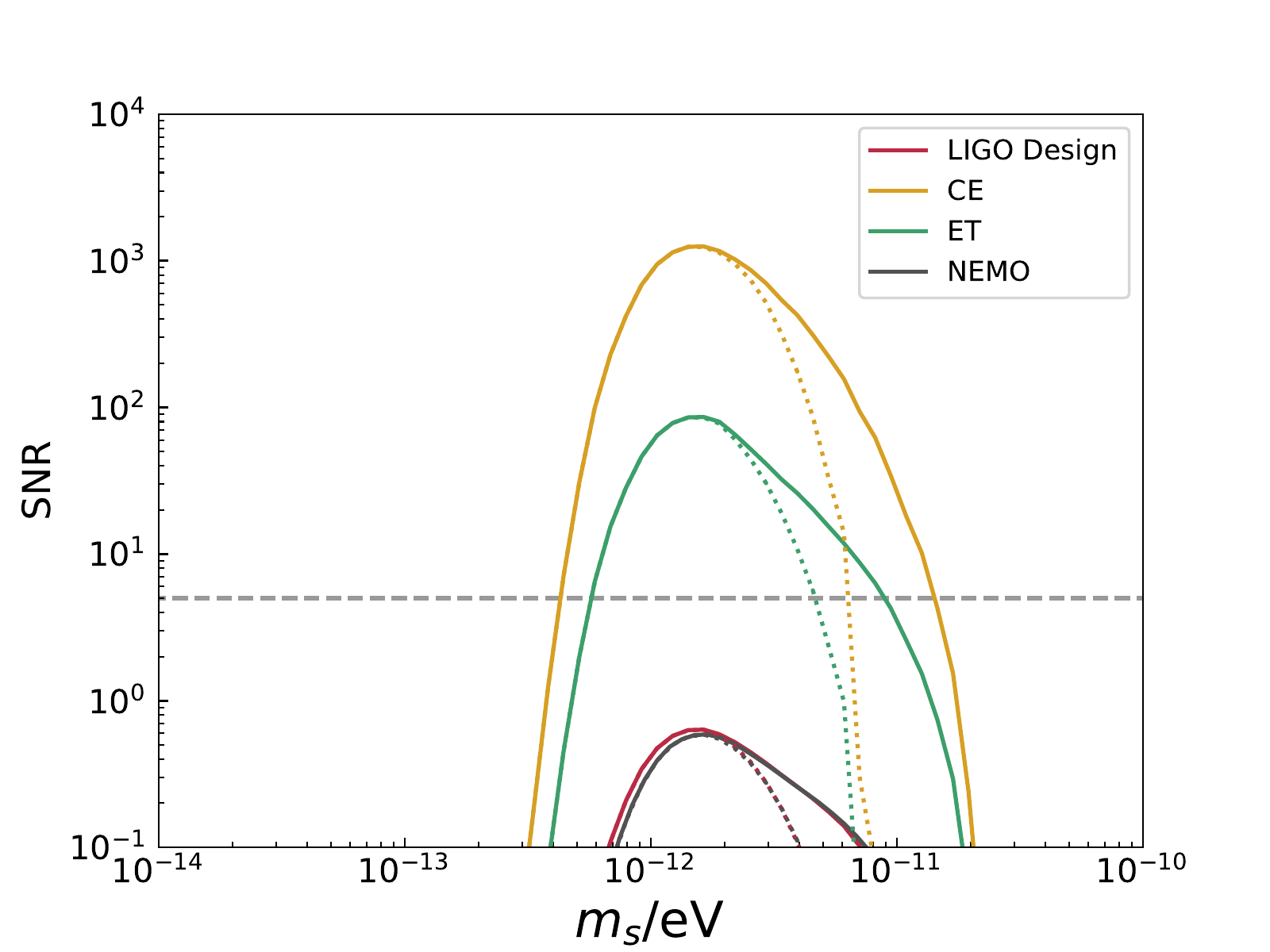}
	\caption{\label{SNRplotPBH}Same as Fig.~\ref{SNRplot} but for the PBH remnant population. Left panel: The mass function of PBHs is $P(M)\propto M^{-3/2}$ in the mass window $M\in [1,5]\Msun$.
		Right panel: The mass function of PBHs has a log-normal form described by Eq.~(\ref{log}). We set $M_p=3\Msun$ and $\sigma=0.5$.
	}
\end{figure*}
Fig.~\ref{SNRplot} shows the SNR of the SGWB from SBHs for the same BH populations considered in Fig.~\ref{highermodes}, whereas Fig.~\ref{SNRplotPBH}  corresponds to the background due to the PBH remnants considered in Fig.~\ref{ogwpbh}. In all cases we highlight the threshold SNR$=5$ below which we expect the SGWB to be too small to be detectable, and compare the case where only the mode $m=1$ is considered (dotted lines) against the case where higher modes are included (solid lines). As one can see, these results confirm that, for $m_s \gtrsim 10^{-12}$ eV, higher modes become dominant and can increase the SNR by more than one order of magnitude compared to the case where only $m=1$ is taken into account. This is especially true for the largest masses and for future detectors.

The SNR can be extremely large confirming previous results that had looked at the LIGO case~\cite{Brito:2017wnc,Tsukada:2018mbp,Tsukada:2020lgt}. In fact, we note that in the range $\sim [10^{-13},10^{-12}]$ eV, the background in the best case scenario is already in tension with upper limits from LIGO observations~\cite{Brito:2017wnc,Tsukada:2018mbp,Tsukada:2020lgt}. For future third-generation (3G) detectors, the SNR could be more than one order of magnitude larger compared to LIGO at design sensitivity. In the best case scenario, observations with CE and ET could probe boson masses in the range $[6.9\times10^{-13},1.7\times 10^{-11}]$ eV and $[9.0\times10^{-13},9.1\times10^{-12}]$ eV, respectively. Interestingly, detectors such as NEMO, that are especially aimed at detecting high frequency GWs in the kHz band, would be competitive with CE and ET for boson masses around $m_s\sim 10^{-11}$ eV.

Importantly, even in the worst-case scenario, where all SBHs are born with negligible spin and only SBH remnants contribute to the background, the background could nonetheless be marginally detectable in LIGO (NEMO) at design sensitivity for bosons with mass $\sim [1.5 , 6.2]\times 10^{-13}$ ($\sim [2.9 , 9.3]\times 10^{-13}$) eV (cf. right panel Fig.~\ref{SNRplot}). The prospects are considerably better with the 3G detectors. Even in this worst-case scenario, the predicted SNR in CE and ET would be orders of magnitude larger and would allow to detect or constrain the boson mass range  $[7.4\times 10^{-14} , 3.2 \times 10^{-12}]$ eV and $[9.1\times 10^{-14} , 2.7 \times 10^{-12}]$ eV, respectively.

Finally, we note that the hypothetical existence of PBH binaries with components masses below $5\, M_{\odot}$ could also lead to a detectable SGWB for bosons with masses up to $10^{-11}$ eV but only with CE or ET, as can be seen in Fig.~\ref{SNRplotPBH}.

\section{Conclusions}
In this paper we explored the capabilities of future ground-based GW detectors to detect the SGWB emitted due to the superradiant instability of ultralight bosons around spinning BHs, extending previous works where this study had been focused on LIGO observations~\cite{Brito:2017wnc,Tsukada:2018mbp,Tsukada:2020lgt}. We paid particular attention to the impact of including unstable modes with $m>1$ in the analysis. Higher modes can dominate the GW emission, especially for bosons with masses $m_s\gtrsim 10^{-12}$ eV. This will be particularly relevant for searches with the future generation of ground-based detectors, which will be able to probe scalar fields in a wide mass window $m_s\in [7\times10^{-14}, 2\times 10^{-11}]$ eV, potentially extending the range that LIGO will be able to probe by almost an one order of magnitude. 

\section*{Acknowledgments}
We acknowledge the use of \texttt{GWSC.jl} package \cite{GWSC} in calculating the sensitivity curves. C.Y. would like to thank Qing-guo Huang and Zu-cheng Chen for useful comments on PBHs and the \texttt{GWSC.jl} package respectively. C.Y. is also indebted to V.C. for his kind hospitality in Lisbon. 
R.B. acknowledges financial support provided under the European Union's H2020 ERC, Starting Grant agreement no.~DarkGRA--757480 and under the MIUR PRIN and FARE programmes (GW-NEXT, CUP:~B84I20000100001). 
V.C. acknowledges financial support provided under the European Union's H2020 ERC 
Consolidator Grant ``Matter and strong-field gravity: New frontiers in Einstein's 
theory'' grant agreement no. MaGRaTh--646597.
This project has received funding from the European Union's Horizon 2020 research and innovation programme under the Marie Sklodowska-Curie grant agreement No 101007855.
We thank FCT for financial support through Project~No.~UIDB/00099/2020.
We acknowledge financial support provided by FCT/Portugal through grants PTDC/MAT-APL/30043/2017 and PTDC/FIS-AST/7002/2020.
The authors would like to acknowledge networking support by the GWverse COST Action 
CA16104, ``Black holes, gravitational waves and fundamental physics.''
We also acknowledge support from the Amaldi Research Center funded by the MIUR program ``Dipartimento di Eccellenza'' (CUP:~B81I18001170001).
	
\bibliography{./ref}
	
\end{document}